\documentstyle[12pt]{article}
\def\fr{\frac}
\def\lbl{\label}
\def\Mck{\frac{M_0^2}{4\k}}
\def\1{\hbox{\bf 1}}
\def\[{\left[}
\def\]{\right]}

\def\gens{generators}
\def\i{\item}
\def\k{\kappa}
\def\s{\sigma}
\def\U{{\cal U}}

\def\ftt{\footnotetext}
\def\ftm{\footnotemark}
\def\bib{\bibitem}
\def\sec{\setcounter{equation}{0}}
\title{The classical basis for $\kappa$-deformed Poincar\'e
(super)algebra
       and the second $\kappa$-deformed supersymmetric Casimir.}

\author{\em P.\ Kosi{\'n}ski \ftm[1]  \ftm[5]  ,
            J.\ Lukierski \ftm[2]  \ftm[6]  ,
            P.\ Ma{\'s}lanka \ftm[3]  \ftm[5],
             and J.\ Sobczyk \ftm[2]  \ftm[6]
}
\date{}

\def\<{\left<}
\def\>{\right>}

\def\e{\epsilon}
\def\ben{\begin{enumerate}}
\def\een{\end{enumerate}}
\def\:{\,\,:\,\,\,}

\def\({\left(}
\def\){\right)}

\def\~{\widetilde}
\def\dsp{\displaystyle}
\def\poin{Poincar\'e }

\def\bowti{\triangleright\!\!\!<}

\newcounter{popnr}
\renewcommand{\theequation}{\arabic{section}.\arabic{equation}}
\def\alpheqn{\setcounter{popnr}{\value{equation}}
             \stepcounter{popnr}
             \setcounter{equation}{0}
             \def\theequation{\arabic{section}.\arabic{popnr}\alph{equation}}
             }
\def\reseteqn{\setcounter{equation}{\value{popnr}}
              \def\theequation{\arabic{section}.\arabic{equation}}
              }
\newcommand{\beq}{\begin{eqnarray}}
\newcommand{\eeq}{\end{eqnarray}}
\newcommand{\beqq}{\begin{eqnarray*}}
\newcommand{\eeqq}{\end{eqnarray*}}

\def\bel#1{\begin{equation}\label{#1}}
\def\be{\begin{equation}}
\def\ee{\end{equation}}
\def\o{\overline}

\def\sP{{\cal P}_{4;1}}
\def\sT{T_{4;2}}

\def\r#1{(\ref{#1})}

\def\P#1{P^{(#1)}}

\def\Q#1{Q^{(#1)}}

\def\bl{\alpheqn}
\def\el{\reseteqn}
\def\a{\alpha}
\def\b{\beta}
\def\ka{{\dot\a}}
\def\kb{{\dot\b}}
\def\ba{\begin{array}}
\def\ea{\end{array}}

\def\epk#1{{\dsp e^{#1\frac{P_0}\k}}}
\def\epkk#1{{\dsp e^{#1\frac{P_0}{2\k}}}}

\def\so{O(3,1;2)}
\def\kdef{$\k$-deformed }
\def\0{^{(0)}}

\begin{document}
\maketitle
\thispagestyle{empty}
\begin{abstract}
We present here the general solution describing generators of \kdef
\poin algebra as the functions of classical \poin algebra generators as well
as the inverse formulae.
Further we present analogous  relations for the generators of
$N=1$ $D=4$ \kdef \poin superalgebra
expressed by the classical \poin superalgebra generators. In such a way we
obtain
the \kdef \poin (super)algebras
 with all the quantum deformation present only in the coalgebra
sector. Using the classical
basis of \kdef \poin superalgebra we obtain as a new result
the $\k$-deformation of supersymmetric covariant spin square
Casimir.
\end{abstract}

\def\thefootnote{\fnsymbol{footnote}}

\ftt[1]{Institute of Physics, University of
{\L}\'od\'z, ul. Pomorska 149/153, 90-236 {\L}\'od\'z, Poland.}
\ftt[2]{
Institute for Theoretical Physics, University of Wroc{\l}aw,
pl. Maxa Borna 9, 50-204 Wroc{\l}aw, Poland.}
\ftt[3]{Dept. of
Functional Analysis, Institute of Mathematics, University of {\L}{\'o}d{\'z},
ul. S. Banacha 22, 90-238  {\L}\'od\'z,
 Poland.}
\ftt[5]{Partially
supported by KBN grant 2P 302 21706.}
\ftt[6]{Partially supported by KBN grant 2P 302 08706.}
\setcounter{footnote}{0}
\def\thefootnote{\arabic{footnote}}

\newpage \setcounter{page}{1}

\sec
\section{Introduction}
The quantum $\k$-deformation with Hopf algebra structure, and mass-like
deformation parameter $\k$ is known for $D=4$ \poin algebra [1--5]
as well as for $D=4$ \poin superalgebra [6--8]. We would like to
point out that recently both the \kdef \poin algebra [5] as well as the
\kdef \poin superalgebra [8] were presented in bicrossproduct basis [9,
10]. In such a basis the algebra is simplified, with free (super)Lorentz
generators and \kdef relations appearing only in the cross-sector of the
(super)Lorentz and (super)translation generators. The aim of this paper
is to simplify further the algebraic sector. In particular
\ben
\item we introduce the basis of \kdef \poin (super)algebra
 with classical (super)\poin generators,
\i we calculate the up to now unknown second Casimir of \kdef $D=4$ \poin
 superalgebra.
\een
In Sect.\ 2 we shall consider the most general formula, expressing the
bicrossproduct basis of \kdef \poin algebra [5] in terms of
classical \poin generators. The only part of the algebra which
undergoes the nonlinear change are the fourmomentum generators, because
the Lorentz generators are already classical in the bicrossproduct
basis. We consider also the inverse formulae, expressing the classical
\poin \gens\ in terms of \kdef ones, and select two special cases of
general
formulas. In Sect.\ 3 we introduce the classical basis for \kdef \poin
superalgebra, in particular for the one given in [8]. It appears that
for such a purpose one should perform in bicrossproduct basis
the nonlinear change of the fourmomenta as well as of two
chiral supercharges. The classical basis of $\k$-\poin superalgebra is
used in Sect.\ 4 for obtaining the \kdef supersymmetric spin Casimir
(supersymmetrized square of Pauli-Lubanski fourvector). With two known
Casimirs of \kdef \poin superalgebra it is possible to develop further
the theory of its irreducible representations, generalizing e.g.\
the results
obtained in [11].
\sec
\section{Classical $\k$-\poin basis}
Let us write the $\k$-\poin algebra as the following cross-product [5]
in its algebraic sector
\bel{2.1}
\U_\k ({\cal P}_4) = \U (O(3,1)) \mathop{\bowti}\limits_{\k\,\,\,}
\U(T_4)\,,
\ee
where  $\U(T_4)$ is the algebra of classical functions of the Abelian
fourmomentum  generators $P_\mu = (P_i,P_0)$ and $\U(O(3,1))$ describes
the enveloping classical Lorentz algebra. The
cross relations between the $O(3,1)$ generators $M_{\mu\nu}=(M_i,N_i)$
and $T_4$ generators $P_\mu=(P_i,P_0)$ are the following [5]:
\bl
\beq
[M_i,P_j]&=&i\e_{ijk}P_k\,,\qquad [M_i,P_0]=0\,, \lbl{2.2a} \\[2mm]
[N_i,P_j]&=& i \delta_{ij} \[ \frac\k2 \(1 - e ^{-\frac{2P_0}\k}\) +
\frac1{2\k} \vec P{}^2\] -\frac{i}\k P_i P_j\,, \lbl{2.2b} \\[2mm]
[N_i,P_0]&=&iP_i\,.\lbl{2.2c}
\eeq
\el

Let us derive the generators of the crossproduct
 basis of $\kappa$-Poincar{\'e}
algebra satisfying \r{2.2a}-\r{2.2c} as functions of classical Poincar{\'e}
generators $(M^{(0)}_i$, $N^{(0)}_i$, $P^{(0)}_i$, $P^{(0)}_0)$. Because in
the
crossproduct basis \r{2.1} the Lorentz algebra is classical we put
\bel{2.3}
\begin{array}{rcl}
M_i&=&M^{(0)}_i, \qquad N_i=N^{(0)}_i\,,\lbl{2.3a}\\[1mm]
P_0&=&f(P^{(0)}_0,M^2_0) \quad P_i=g(P^{(0)}_0,M^2_0)P^{(0)}_i
\end{array}
\ee
where
\bel{2.4}
M^2_0=(P^{(0)})^2-(\vec P{}^{(0)})^2
\ee
is the mass Casimir of the classical Poincar{\'e} algebra. One obtains
from the substitution of the relations \r{2.3} in
the relations \r{2.2a}-\r{2.2c} that $(f'\equiv
\frac{\partial f}{\partial P^{(0)}_0})$
\bel{2.5}
g'=-\frac{1}{\kappa}g^2\,, \qquad f'=g\,.
\ee
The general solutions of equation \r{2.5} are the following
\bel{2.6}
g=\frac{\kappa}{\P0_0+C}\,,\qquad
f=\kappa\ln\frac{\P0_0+C}{A}\,,
\ee
where from the relations
\bel{2.7}
\lim\limits_{\kappa\longrightarrow\infty}P_i=\P0_i\,, \qquad
\lim\limits_{\kappa\longrightarrow\infty}P_0=\P0_0\,,
\ee
one gets
\bel{2.8}
\lim\limits_{\kappa\longrightarrow\infty}\frac{A}{\kappa}=
\lim\limits_{\kappa\longrightarrow\infty}
\frac{C}{\kappa}=1\,.
\ee
Besides one obtains the relation
\bel{2.9}
gP^{(0)}_0=\frac{\kappa}{2}(1-e^{-2\frac{P_0}{\kappa}})+
\frac{1}{2\kappa}P^2_i\,,
\ee
which implies that
\bel{2.10}
C^2-A^2=M^2_0\,.
\ee
We see therefore that one obtains the family of solutions depending on
one arbitrary function $A(\kappa ;M^2_0)$, satisfying the condition
\r{2.8}.

The \kdef Casimir in Majid-Ruegg basis takes the form
\bel{2.11}
\(2\k \sinh \frac{P_0}{2\k} \)^2 - \epk{} P_i^2 =M^2\,.
\ee
Substituting the formulae \r{2.3} one gets
\bel{2.12}
M^2=2\k^2\( \frac CA -1\)\,.
\ee
Using \r{2.10} one obtains the following relations between two Casimirs
\r{2.10} and \r{2.11}
\bel{2.13}
M_0^2=A^2 \[ \( \frac{M^2}{2\k^2} +1\)^2-1\]\,.
\ee
The general inverse formulae look as follows:
\bl
\beq
\P0_0&=&\dsp A \({\dsp\epk{}} -\frac CA\) =\dsp A\ \({\dsp \epk{}} -1 -
\frac {M^2}{2\k^2}
\)\,, \lbl{2.14a}\\[2mm]
\P0_i&=&\dsp \frac A\k \epk{} P_i\,. \lbl{2.14b}
\eeq
\el
We shall consider the following two particular solutions:
\ben
\i[a)]
\bel{2.15}
A=\k - \Mck \,,\qquad C= \k + \Mck \,.
\ee
We get
\bel{2.16}
\ba{rcl}
P_0&=&\dsp \k \ln \( \frac{\P0_0 +\k + \Mck}{\k- \Mck}\)\,,\\[3mm]
P_i&=&\dsp \frac{\k}{\P0_0+\k+\Mck} \P0_j
\ea
\ee
and the inverse formulae
\bel{2.17}
\ba{rcl}
\P0_0 &=&\dsp \k \( \epk{} -1 \) - \Mck\(\epk{}+1\)\,,\\[2mm]
\P0_i &=&\dsp \( 1- \Mck\) P_i \epk{}
\ea
\ee
where in \r{2.17} one should use the formulae \r{2.13}
giving for the choice \r{2.15} the result
\bel{2.18}
M^2_0 =\fr{M^2}{1+\fr{M^2}{4\k^2}}\,,
\ee
\i[b)]
\bel{2.19}
A=\k\,,\qquad C=\(M_0^2+\k^2\)^{\frac12}\,.
\ee
One obtains
\bel{2.20}
\ba{rcl}
P_0&=&\dsp \k\ln \[ \P0_0+\(M_0^2+\k^2\)^{\frac12}\]-\k\ln\k\,,\\[2mm]
P_i&=&\dsp \frac{\k}{\P0_0+\(M_0^2+\k^2\)^{\frac12}} \P0_i
\ea
\ee
and
\bel{2.21}
\ba{rcl}
\P0_0&=&\dsp \k\sinh\ \frac{P_0}\k + \frac{1}{2\k} \epk{}
{\vec P{}}^2\,,\\[2mm]
\P0_i&=&\dsp e^{\frac{P_0}\k} P_i\,.
\ea
\ee
The formula \r{2.13} takes the form
\bel{2.22}
M_0^2 = {M^2}\({1+\fr{M^2}{4\k^2}}\)\,.
\ee
\een

It should be mentioned  that the formulae \r{2.21}-\r{2.22} have been
already obtained in [12], and it has been observed by Ruegg [13] that the
formulae \r{2.21} applied to bicrossproduct basis provide the classical
\poin algebra.

In the basis given by \r{2.14a}-\r{2.14b} (see also \r{2.17} and \r{2.21})
all the
quantum deformation is contained in the crossproduct.
The relations are quite complicated and the coproduct for the energy
operator $\P0_0$ does not lead to the additivity of the energy.

\sec
\section{Classical $\k$-super\poin basis}

Recently it has been shown [8] that the Majid-Ruegg basis with the
cross-relations \r{2.2a}-\r{2.2c} can be supersymmetrized. The relation
\r{2.1} is supersymmetrized as follows \footnote{We discuss here only
one explicite way of describing the \kdef \poin superalgebra as a graded
bicrossproduct. For other ways see [8].}:
\bel{3.1}
\U_\k(\sP)=\U(\so) \bowti \U(\o \sT)\,,
\ee
where $\U(\o \sT)$ is the graded algebra of functions on the ``trivial''
chiral superalgebra:
\bel{3.2}
\o \sT \: \{ \o Q _\ka, \o Q_\kb\} = [P_\mu,\o Q_\ka] = [P_\mu,P_\nu] =0
\ee
and $\U(\so)$ describes the enveloping algebra of the graded Lorentz
algebra $\so=(M_{\mu\nu}, Q_\a)$, with the Lorentz algebra supplemented
by the relations
\bl
\beq
[M_{\mu\nu},Q_\a] &=&\frac12 (\s_{\mu\nu})_\a{}^\b Q_\b\,,\lbl{3.3a}\\[2mm]
\{Q_\a,Q_\b\}&=&0\,.\lbl{3.3.b}
\eeq
\el
The cross-relations in \r{3.1} are given by the following extension of
the set of relations \r{2.2a}-\r{2.2c} ($M_{\mu\nu}\equiv(
M_i=\frac12\e_{ijk}M_{jk}$,
$N_i=M_{0i})$):
\bl
\beq
[M_i,\o Q_\ka] &=&\frac12 ( \o Q \s_i)_\ka \,,\lbl{3.4a}\\[2mm]
[N_i, \o Q_\ka]&=& -\frac i2 \epk- (\o Q \s_i)_\ka +
\frac1{2\k}\e_{ijk}P_j(\o Q \s_k)_\ka\,,\lbl{3.4b}\\[2mm]
[Q_\a,P_\mu]&=& 0\,,\lbl{3.4c} \\[2mm]
\{ Q_\a,\o Q_\kb\} &=& 4\k \delta_{\a\kb} \sinh\frac{P_0}{2\k} - 2
P_i(\s_i)_{\a\kb}\epkk{} \,.\lbl{3.4d}
\eeq
\el
It is known from Sect.\ 2 that the fourmomentum operators $P_\mu$
satisfying \r{2.2a}-\r{2.2c} can be expressed as functions of the
fourmomenta $\P0_\mu$ belonging to the classical \poin algebra.
Similarly, one can express the generators $Q_\a$, $\o Q_\ka$
in terms of classical
$N=1$ \poin superalgebra generators ($ \o Q^{(0)}_\ka, \P0_\mu$),
satisfying the relations
\bel{3.5}
\ba{rcl}
\{\Q0_\a,\o \Q0_\kb\}&=& 2 \delta _{\a\kb}\P0_0 - 2
(\s_i)_{\a\kb}\P0_i\,,\\[2mm]
\{\Q0_\a, \Q0_\b\} &=& \{\o \Q0_\ka,\o \Q0_\kb\}\, =\, 0\,,\\[2mm]
[\Q0_\a, P_\mu ]& =& [ \o \Q0_\ka,P_\mu ]\, =\,0\,.
\ea
\ee
We obtain that
\bl
\beq
Q_\a&=&\dsp\sqrt{\frac{2\k^2}{A(A+C)}} \Q0_\a\,,\lbl{3.6a} \\[2mm]
\o Q_\ka &=&\dsp  \o \Q0_\kb [\exp (\a \frac{\P0_k}{|\vec P{}^{(0)}|}
 \s_k)]^\kb{}_\ka\,,\lbl{3.6b}
\eeq
\el
where
\bl
\beq
\sinh \a &=& \frac{\left|\vec
P{}^{(0)}\right|}{[2(\P0_0+C)(A+C)]^\frac12}\,,\lbl{3.7a}\\[2mm]
\cosh \a &=& \frac{\P0_0+C+A}{[2(\P0_0+C)(A+C)]^\frac12}\,.\lbl{3.7b}
\eeq
\el
It is easy to check that $\cosh^2\a-\sinh^2\a=1$ follows
from the relation \r{2.10}.

In order to obtain the formulae expressing the classical supercharges
$\Q0_\a$, $\o \Q0_\ka$ in terms of \kdef generators  $Q_\a$, $\o Q_\ka$,
$P_\mu$ one should substitute in \r{3.6b} the relations
\r{2.14a}-\r{2.14b} and put
the exponential on the other side of the equation \r{3.6b}.

\sec
\section{The second Casimir for the $N=1$ $\kappa$-Poincar{\'e}
superalgebra}

\setcounter{equation}{0}

In [6] it was observed that the $\kappa$-deformed mass Casimir \r{2.11} of
the $\kappa$-Poincar{\'e} algebra  becomes
the Casimir of $\kappa$-deformed $N=1$ superPoincar{\'e} algebra. The
second relativistic spin square Casimir of the $\kappa$-Poincar{\'e}
algebra takes the form [2, 3]
\bel{4.1}
C_2=(\cosh\frac{P_0}{\kappa}-\frac{{\vec P}{}^2}{4\kappa^2})W^2_0-{\vec
W}{}^2
\ee
where
\setcounter{popnr}{\value{equation}}
\addtocounter{popnr}{1}
\bl
\beq
W_0&=&{\vec P\cdot\vec M}\,,\lbl{4.2a}\\[2mm]
W_k&=&\kappa M_k\sinh\frac{P_0}{\kappa}+\epsilon_{ijk}P_iL_j
\eeq
\el
We construct the supersymmetric extension of the $\kappa$-deformed
Casimir $C_2$ as follows:
\begin{description}
\item[i)] Using the classical superalgebra basis of $N=1$
$\kappa$-superPoincar{\'e} algebra (see Sect.3) the supersymmetric
extension $C_2^{N=1}$ of the relativistic spin square Casimir is
described by the length of the supersymmetric Pauli-Lubanski vector
\bel{4.3}
C_2^{N=1} =W_{\mu}^{N=1}W^{\mu N=1}\equiv (W_0^{N=1})^2 -{\vec W}{}^{N=1}
{\vec W}{}^{N=1}
\ee
where [14]
\beq
W^{N=1}_0&=&{\vec P}{}^{(0)}{\vec M}+T^{(0)}_0\,,\lbl{4.4} \\[2mm]
W_i^{N=1}&=&M_iP^{(0)}_0+\epsilon_{kij}P^{(0)}_iN_j^{(0)}
            +T^{(0)}_i \lbl{4.5}
\eeq
and (see [6])
\bel{4.6n}
T^{(0)}_{\mu}= \Q0_\a \s_{\mu}{}^{\a\kb}\o \Q0 _\kb\,.
\ee
\item[ii)] In the crossproduct basis given in [8] (see also Sect. 3)
the second Casimir
\r{4.3} can be expressed in terms of the \kdef generators provided that
the formulae \r{2.14a}-\r{2.14b} and \r{3.6a}-\r{3.6b} are properly employed.
One gets
\bl
\beq
T^{(0)}_0&=&\dsp \frac{A}{\kappa}
\cosh{\frac{P_0}{2\kappa}}\,T_0+\frac{A}{2\kappa^2}e^{\frac{P_0}{2\kappa}}
P_jT_j\,,\lbl{4.6a}\\[1mm]
T^{(0)}_j&=&
\begin{array}[t]{l}\dsp
\frac{A}{\kappa}\cosh\frac{P_0}{2\kappa}T_j+
\frac{A}{2\kappa^2}e^{\frac{P_0}{2\kappa}}P_jT_0-
\\[1mm]\dsp
\null-i\frac{A}{2\kappa^2}\epsilon_{jkl}e^{\frac{P_0}{2\kappa}}P_kT_l\,,
\lbl{4.6b}
\end{array}
\eeq
\el
where $T_\mu=(T_i,T_0)$ is given by the formulae \r{3.6a}-\r{3.6b}
with the supercharges
$\Q0_\a$, $\o \Q0_\ka$ replaced with $Q_\a$, $\o Q_\ka$.

One obtains
\alpheqn
\beq
W_0^{N=1}&=&\begin{array}[t]{l}\dsp
\frac{A}{\kappa}\lbrack
e^{\frac{P_0}{\kappa}}P_jM_j+\cosh
(\frac{P_0}{2\kappa})T_0+\null\\[2mm]\dsp
\null+\frac{1}{2\kappa}\epkk{} P_kT_k
\rbrack\,,
\end{array}\lbl{4.7a} \\[2mm]
W^{N=1}_i&=&
\begin{array}[t]{l}
\dsp \frac{A}{\kappa}\lbrack M_j (\kappa
e^{\frac{P_0}{\kappa}}-\kappa\frac{C}{A})+\epsilon_{ikl}
e^{\frac{P_0}{\kappa}}P_kN_l+\\[2mm]\dsp
+\cosh
(\frac{P_0}{2\kappa})T_j+\frac{1}{2\kappa}
e^{\frac{P_0}{2\kappa}}P_jT_0-\null\\[2mm]\dsp
-\frac{i}{2\kappa}\epsilon_{ikl}e^{\frac{P_0}{2\kappa}}P_kT_l
\rbrack\,.
\end{array}         \lbl{4.7b}
\eeq
\el
The relations \r{4.7a}-\r{4.7b} should be subsequently substituted in
\r{4.3}
\item[iii)] In the conventional basis given in [6]
 the second Casimir \r{4.3} is obtained if we substitute in
 \r{4.7a}-\r{4.7b}
  the formulae relating standard basis (see [6]) with the bicrossproduct
  basis (see [8]).
\end{description}

\section{Final Remarks}

In this note we discussed the $\k$-\poin (super)algebra with the basis
described by the generators satisfying classical (super-)\poin algebra.
We recall that in such
a basis all the deformation is present in the coalgebra sector.
We would also like to mention that for any choice of the
fourmomenta in classical basis
(see \r{2.17}) the energy ceases to be additive.

It is known that the coalgebra sector of the quantum Lie algebra
determines the algebraic relations satisfied by the generators of dual
quantum Lie group. For the bicrossproduct basis of $\k$-\poin algebra
[5] it was recently shown [15] that the quantum \poin group is described
by the relations firstly obtained by Zakrzewski [16] via quantization of
the Lie-Poisson structure. It would be very interesting to generalize this
result to the case of \kdef \poin superalgebra for which the
super-extension of the quantum \poin group of Zakrzewski was given in
[7]. Another problem which is interesting to solve is to provide a
basis of the quantum \poin group dual to the classical
basis \r{2.17} of the $\k$-\poin algebra.
Because the coproduct for $\k$-\poin algebra with classical
basis is complicated, one should expect complicated algebraic relation
for these \poin group generators.

The question which remains still not well understood in the framework of
\kdef symmetry is the physical meaning of \kdef coproducts. One can say
only very generally that it describes some kind of basic interaction
dictated by quantum deformation of underlying relativistic symmetry.

\end{document}